\documentclass[final, 3p, review,11pt]{elsarticle}
\usepackage{natbib}
\usepackage{graphicx}
\usepackage{color}
\usepackage{amsmath}

\newlength\figwidth
\journal{Elsevier Science}

\begin{document}

\begin{frontmatter}

\title{Parametric polynomial minimal surfaces of arbitrary degree}

\author[A]{Gang Xu} \ead{gxu@sophia.inria.fr}
\author[B]{Guozhao Wang}
 
\address[A]{Galaad, INRIA Sophia-Antipolis, 2004 Route des Lucioles, 06902 Cedex, France}
\address[B]{Department of Mathematics, Zhejiang University, Hangzhou, China}

\begin{abstract}

Weierstrass representation is a classical parameterization of minimal
surfaces. However, two functions should be specified to construct 
the parametric form in Weierestrass representation. In this paper,
we propose an explicit parametric form for a class of parametric polynomial minimal surfaces of
arbitrary degree. It includes the classical Enneper surface for cubic case. 
The proposed minimal surfaces also have some interesting properties
such as symmetry, containing straight lines and self-intersections. 
According to the shape properties, the proposed minimal surface can be
classified into four categories with respect to $n=4k-1$ $n=4k+1$,
$n=4k$ and $n=4k+2$. The explicit parametric form of corresponding 
conjugate minimal surfaces is given and the isometric deformation 
is also implemented.

\end{abstract}


\end{frontmatter}

\section{Introduction}
Minimal surface is a kind of surface with vanishing mean
curvature \cite{Nitsche:lecture}. As the mean curvature is the variation of area
functional, minimal surfaces include the surfaces minimizing the
area with a fixed boundary \cite{Morgan1, Morgan2}.
There have been many literatures on minimal surface in classical
differential geometry \cite{Osserman, Meeks1, Meeks2}. Because of their attractive properties,
minimal surfaces have been extensively employed in many areas such
as architecture, material science, aviation, ship manufacture,
biology and so on. For instance, the shape of the membrane
structure, which has appeared frequently in modern architecture, is
mainly based on minimal surfaces \cite{Blet:minimal}. Furthermore,
triply periodic minimal surfaces naturally arise in a variety of
systems, including nano-composites, lipid-water systems and certain
cell membranes \cite{Jung:levelset}.

In CAD systems, parametric polynomial representation is the standard
form. For parametric polynomial minimal surface, plane is the unique
quadratic parametric polynomial minimal surface, Enneper surface is
the unique cubic parametric polynomial minimal
surface.  There are few research work on
the parametric form of polynomial minimal surface with higher
degree. Weierstrass representation is a classical parameterization of minimal
surfaces. However, two functions should be specified to construct 
the parametric form in Weierestrass representation. In this paper,
we discuss the answer to the following questions: What are 
the possible explicit parametric form of polynomial minimal surface of
arbitrary degree and how about their properties?
The proposed minimal surfaces include the classical Enneper surface
for cubic case, and also have some interesting properties
such as symmetry, containing straight lines and self-intersections. 
According to the shape properties, the proposed minimal surface can be
classified into four categories with respect to $n=4k-1$ $n=4k$,
$n=4k+1$ and $n=4k+2$. The explicit parametric form of corresponding 
conjugate minimal surfaces is given and the isometric deformation 
is also implemented.  

The paper includes five sections. \emph{Preliminary}
introduces some notations and lemmas. \emph{Main  Results} presents the 
explicit parametric formula of parametric polynomial minimal surface
of arbitrary degree. The next section, \emph{Properties and
  Classification} presents the corresponding properties and
classification of the proposed minimal surfaces. The following section
focuses on corresponding conjugate counterpart of proposed minimal
surface. Finally, in \emph{Conclusions}, we summarize the main
results. 
\section{Preliminary}

In this section, we introduce the following two notations.
\begin{eqnarray}
P_n & = & \sum\limits_{k=0}^{\lceil \frac{n-1}{2} \rceil}(-1)^k{n \choose
 2k}u^{n-2k}v^{2k},\label{Pn} \\
Q_n & = & \sum\limits_{k=0}^{\lfloor \frac{n-1}{2} \rfloor}(-1)^k{n \choose
 2k+1}u^{n-2k-1}v^{2k+1}\label{Qn}
\end{eqnarray}

$P_n$ and $Q_n$ have the following properties:

\noindent\textbf{Lemma 1 }\label{lemma:deriv}
 \begin{eqnarray*}
&&\frac{\partial P_n}{\partial u } =  n P_{n-1}, \quad 
 \frac{\partial P_n}{\partial v } =  -n Q_{n-1} \\
&&\frac{\partial Q_n}{\partial u } = n Q_{n-1}, \quad
 \frac{\partial Q_n}{\partial v } =  n P_{n-1}
\end{eqnarray*}

\noindent\textbf{Lemma 2 }\label{lemma:iteration}
  \begin{eqnarray*}
    P_{n}& = & uP_{n-1}-vQ_{n-1}  \\
    Q_{n}& = & vP_{n-1}+uQ_{n-1}  \\
  \end{eqnarray*}
Lemma \ref{lemma:iteration} can be proved by using the following equation: 
\[
{n \choose 2k}+{n \choose 2k+1}={n+1 \choose 2k+1}. 
\]

\section{Main Results}

\noindent\textbf{Theorem 1 }\label{ref:form}
If the parametric representation
of polynomial surface $\textbf{r}(u,v)$ with arbitrary
degree $n$, is given by $\textbf{r}(u,v)=(X(u,v), Y(u,v), Z(u,v))$,
where
\begin{eqnarray} \label{eq:definition}
X(u,v) & = & -P_n+\omega P_{n-2},\nonumber\\
Y(u,v) & = & Q_n+\omega Q_{n-2},\label{formula}\\
Z(u,v) & = & \frac{ 2\sqrt {n(n-2)\omega}}{
n-1} P_{n-1}, \nonumber
\end{eqnarray}
then $\textbf{r}(u,v)$ is a minimal surface.

\noindent\textbf{Proof of Theorem 1.} From  Lemma \ref{lemma:deriv}, we have 
\begin{eqnarray*}
 \frac{\partial^2 \emph{\textbf{r}}(u,v)}{\partial^2 u}+\frac{\partial^2
\emph{\textbf{r}}(u,v)}{\partial^2 v} =0
\end{eqnarray*}
Hence, $\emph{\textbf{r}}(u,v)$ is harmonic surface. 

By using Lemma \ref{lemma:deriv}, we have
\begin{eqnarray}
F&=&\frac{\partial \emph{\textbf{r}}(u,v)}{\partial u}\frac{\partial
  \emph{\textbf{r}}(u,v)}{\partial v} \\
 &=& 2n(n-2)\omega( Q_{n-3}P_{n-1}+P_{n-3}Q_{n-1}-2Q_{n-2}P_{n-2}) \nonumber
\end{eqnarray} 
From Lemma \ref{lemma:iteration},  we have
\begin{eqnarray}
P_{n-2}&=& uP_{n-3}-vQ_{n-3},\\
Q_{n-2}&=& vP_{n-3}+uQ_{n-3},\\
P_{n-1}&=& (u^2-v^2)P_{n-3}-2uv Q_{n-3},\\
Q_{n-1}&=& (u^2-v^2)Q_{n-3}+2uv P_{n-3},
\end{eqnarray}
Substituting (5)(6)(7)(8) into (4), we can obtain $F=0$. 
Similarly, we have
\begin{eqnarray*}
  E-G&=&\frac{\partial \emph{\textbf{r}}(u,v)}{\partial u} \frac{\partial \emph{\textbf{r}}(u,v)}{\partial u}-\frac{\partial \emph{\textbf{r}}(u,v)}{\partial v} \frac{\partial \emph{\textbf{r}}(u,v)}{\partial v}\\
  &=& 4n(n-2)\omega( Q_{n-1}Q_{n-3}-P_{n-3}P_{n-1}+P^2_{n-2}-Q^2_{n-2})\\
  &=&0
\end{eqnarray*} 
Hence,  $\emph{\textbf{r}(u,v)}$ is 
a parametric surface with isothermal parameterization.  
From \cite{Nitsche:lecture}, if a parametric surface with isothermal parameterization  is 
harmonic, then it is a  minimal surface. The proof is
completed.

\section{Properties and Classification}
From \emph{Theorem 1}, if $n=3$, we can get the Enneper surface,
which is the unique cubic parametric polynomial minimal surface. It has the following parametric form
\begin{equation*}
\textbf{\emph{E}}(u,v)=(-(u^3-3uv^2)+\omega u, -(v^3-3vu^2)+\omega
v, \sqrt{3\omega}(u^2-v^2)).
\end{equation*}
Enneper surface has several  interesting properties, such as symmetry,
self-intersection, and containing orthogonal straight lines on it. 
For the new proposed minimal surface, we can also prove that it has
also has these properties. 

If $n=5$, a kind of quintic polynomial minimal surface proposed in \cite{Gang1} can be obtained
as follows
\begin{equation}
 \textbf{\emph{Q}}(u,v)=(X(u,v), Y(u,v), Z(u,v)).
 \end{equation}
where
\begin{eqnarray*}
X(u,v) & = & -(u^5-10u^3v^2+5uv^4)+\omega u(u^2-3v^2),\\
Y(u,v) & = & -(v^5-10v^3u^2+5vu^4)+\omega v(v^2-3u^2),\\
Z(u,v) & = & \frac{\sqrt{15\omega}}{2}(u^4-6u^2v^2+v^4).
\end{eqnarray*}
According to the shape properties, the proposed minimal surface in
\emph{Theorem 1} can be classified into four
classes with $n=4k-1$ $n=4k$, $n=4k+1$, $n=4k+2$.  

\noindent\textbf{Proposition 1. }  In case of $n=4k-1$, the
corresponding proposed minimal surface $\emph{\textbf{r}}(u,v)$ has the following properties:
\begin{itemize}
\item $\emph{\textbf{r}}(u,v)$ is symmetric about the plane $X=0$ and  the plane $Y=0$,
\item $\emph{\textbf{r}}(u,v)$ contains two orthogonal straight lines $x=\pm y$ on the plane $Z=0$
\end {itemize} 
 
Fig. 1(a) shows an example of Enneper surface,
Fig. 1 (b) shows an example of
proposed minimal surface with $n=7$. The symmetry plane and straight
lines of minimal surface in Fig. 1 (b) are shown in
Fig. 1 (c) and Fig. 1 (d). 

\noindent\textbf{Proposition 2. }  In case of $n=4k$, the
corresponding proposed minimal surface $\emph{\textbf{r}}(u,v)$ is
symmetric about the plane $Z=0$ and  the plane $Y=0$.

\begin{figure}[t]\label{fig:cubic}
\centering
\begin{minipage}[t]{0.6\figwidth}
\centering
\includegraphics[height=1.0\textwidth,width=1.1\textwidth]{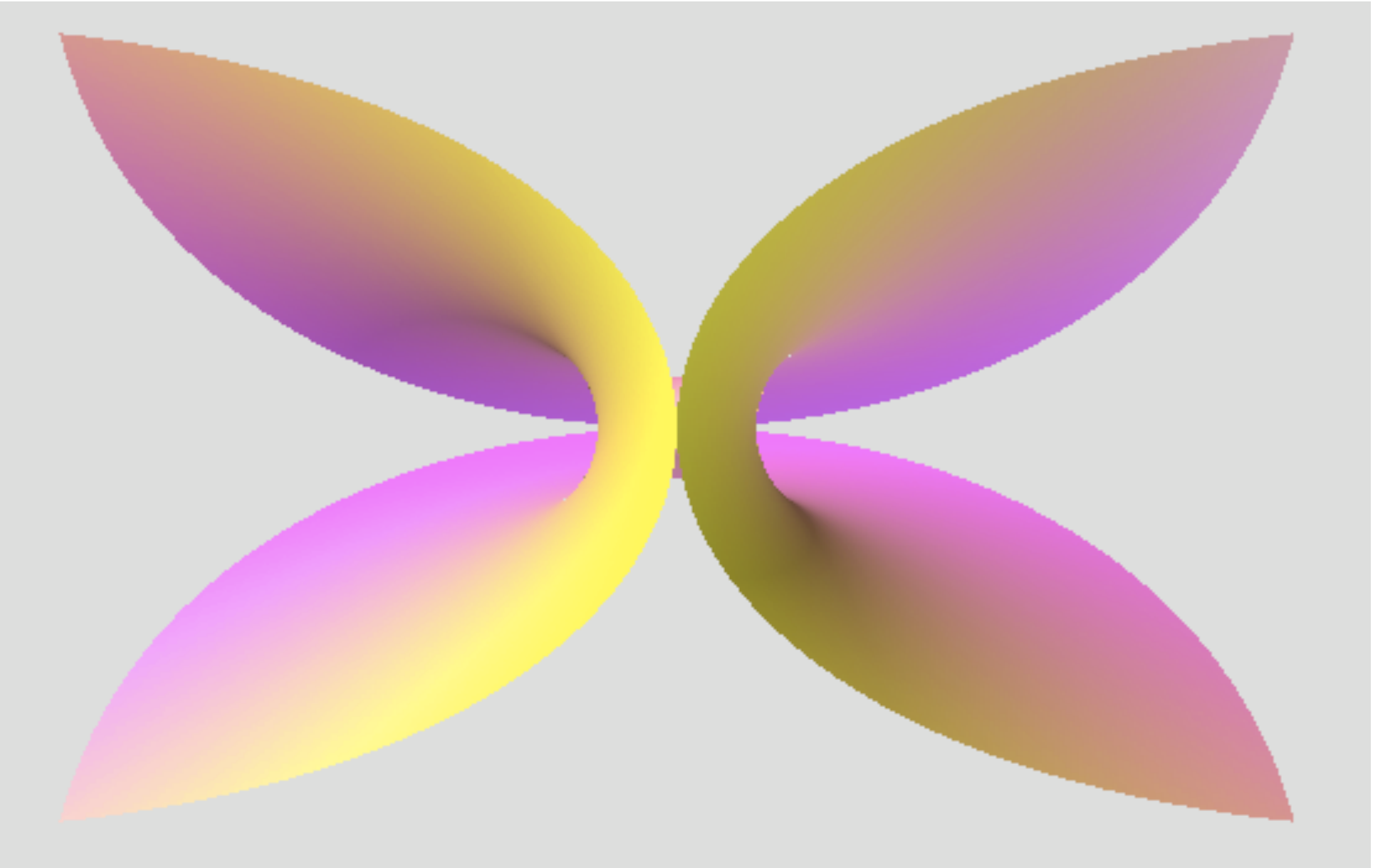}
\\(a) Ennerper surface 
\end{minipage} \quad
\begin{minipage}[t]{0.6\figwidth}
\centering
\includegraphics[height=1.0\textwidth,width=1.1\textwidth]{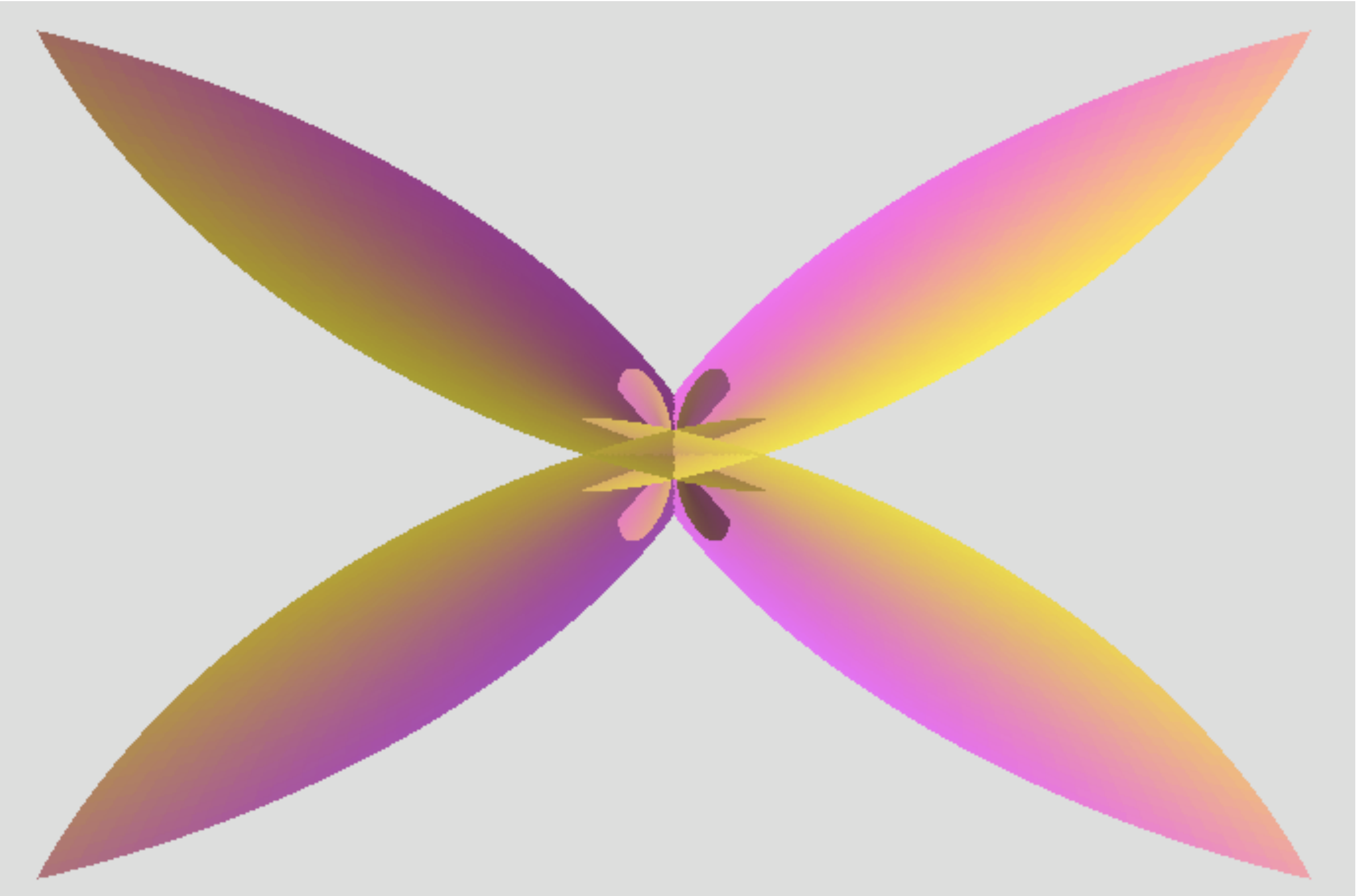}
\\(b) Polynomial surface of degree 7
\end{minipage}\\
\begin{minipage}[t]{0.6\figwidth}
\centering
\includegraphics[height=1.0\textwidth,width=1.1\textwidth]{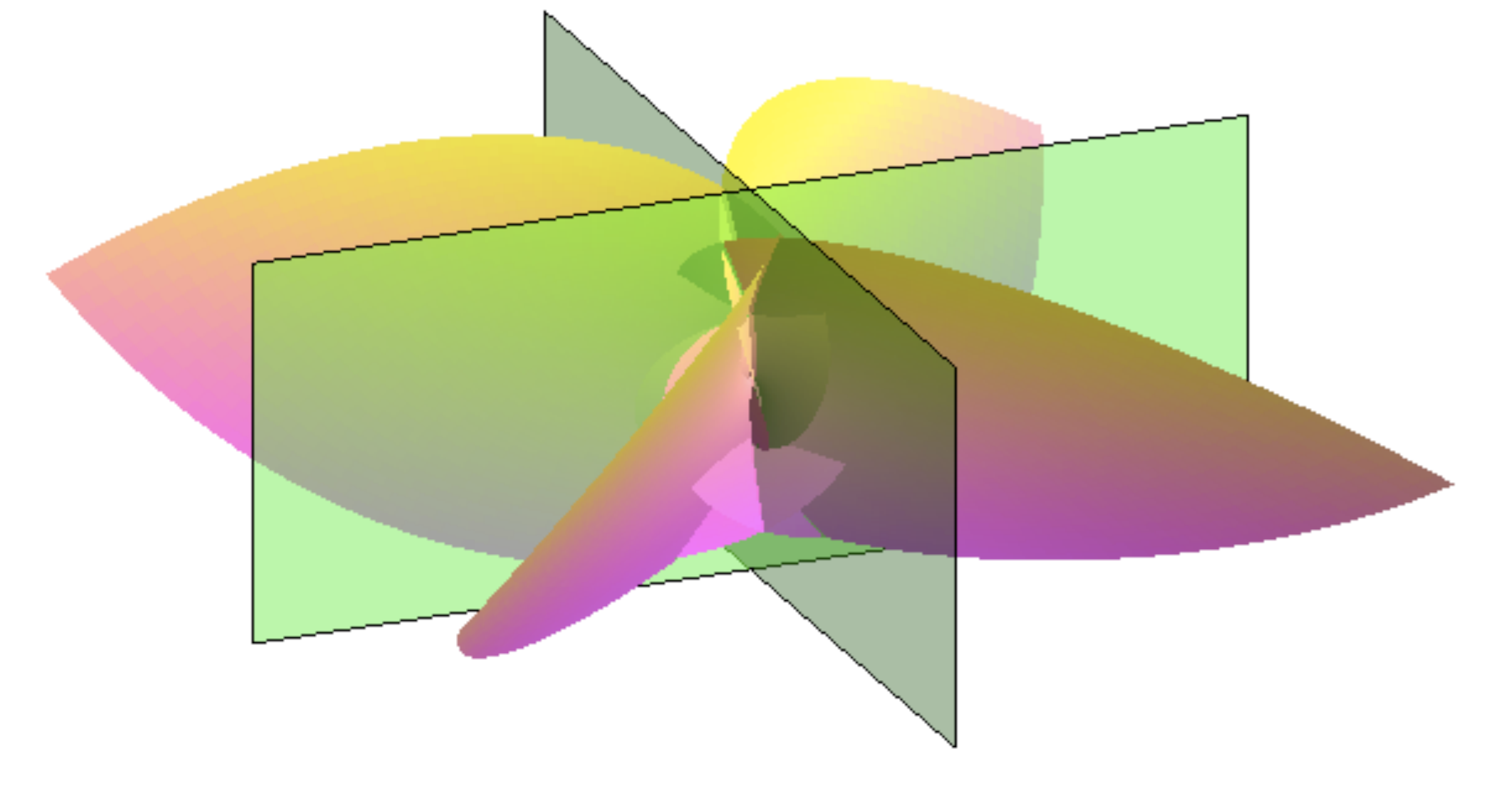}
\\(a) Symmetry plane
\end{minipage} \quad
\begin{minipage}[t]{0.6\figwidth}
\centering
\includegraphics[height=1.0\textwidth,width=1.1\textwidth]{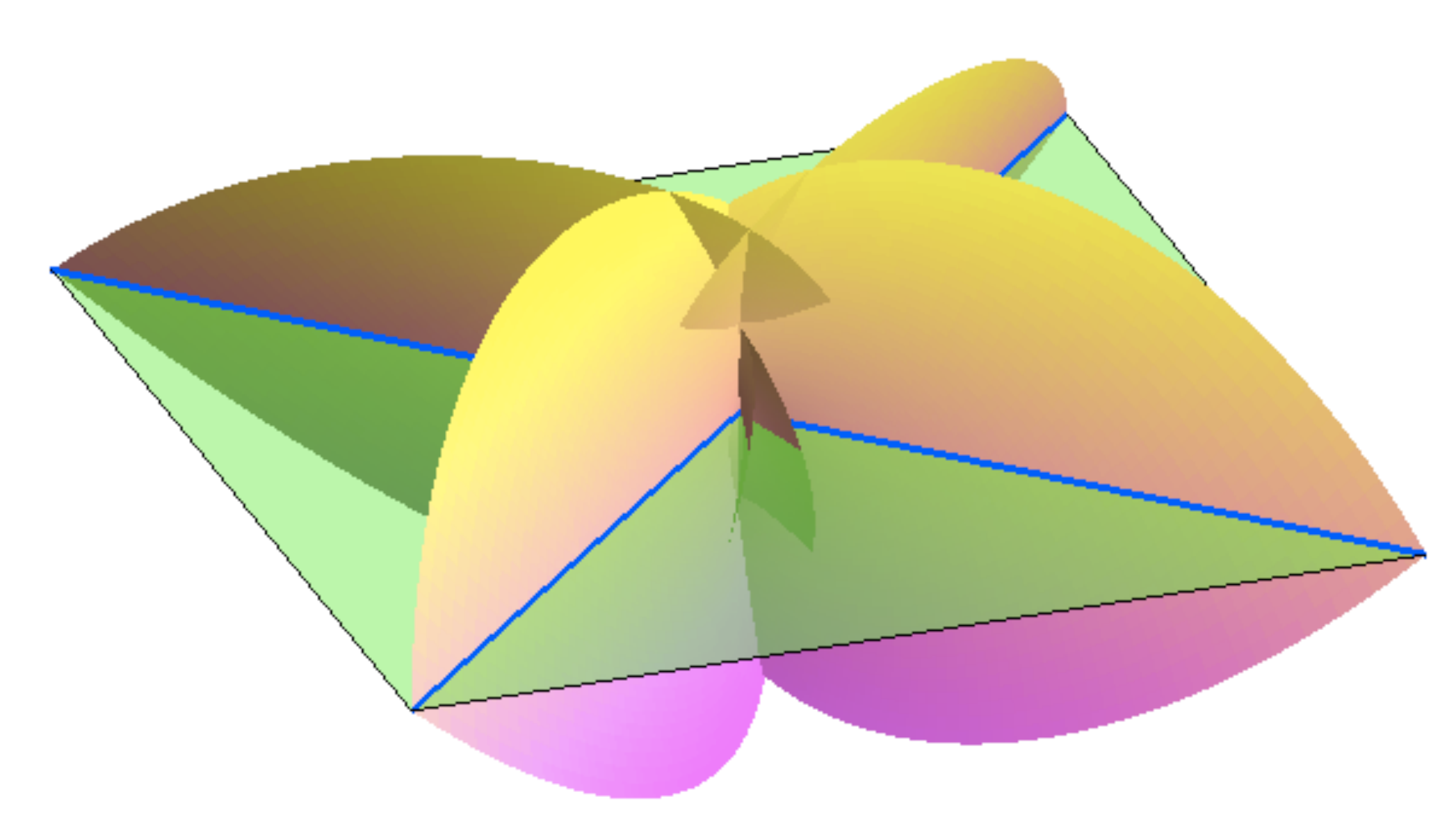}
\\(b) Straight lines 
\end{minipage}
\caption{Enneper surface and minimal surface of degree seven.  $\omega=1, [-1,1]\times[-1,1]$.}
\end{figure}

Fig. 2 (a)  present an example of 
proposed quartic minimal surface and the corresponding symmetry planes are shown in
Fig. 2 (b).

\noindent\textbf{Proposition 3. }  In case of $n=4k+1$, the
corresponding proposed minimal surface $\emph{\textbf{r}}(u,v)$ has the following properties:
\begin{itemize}
\item $\emph{\textbf{r}}(u,v)$ is symmetric about the plane $X=0$,  the plane $Y=0$,
  the plane  $X=Y$ and the plane $X=-Y$. 
\item Self-intersection points of $\emph{\textbf{r}}(u,v)$ are only on the symmetric planes, i.e.,
there are no other self-intersection points on
$\underline{\emph{\textbf{r}}}(u,v)$, and the self-intersection
curve has the same symmetric plane as the minimal surface.
\end {itemize} 

Fig. 3 (a)  present an example of 
proposed quintic minimal surface  and the corresponding symmetry planes are shown in
Fig. 3 (b).

\noindent\textbf{Proposition 4. }  In case of $n=4k+2$, the
corresponding proposed minimal surface $\emph{\textbf{r}}(u,v)$ is
symmetric about the plane $Z=0$ and  the plane $Y=0$.

For the case of $n=6$, it has been studied in \cite{Gang2}. Fig.4 (a)  present an example of 
proposed minimal surface with $n=6$ and the corresponding symmetry planes are shown in
Fig. 4 (b).

\begin{figure}[t]\label{fig:cubic}
\centering
\begin{minipage}[t]{0.6\figwidth}
\centering
\includegraphics[height=1.0\textwidth,width=1.1\textwidth]{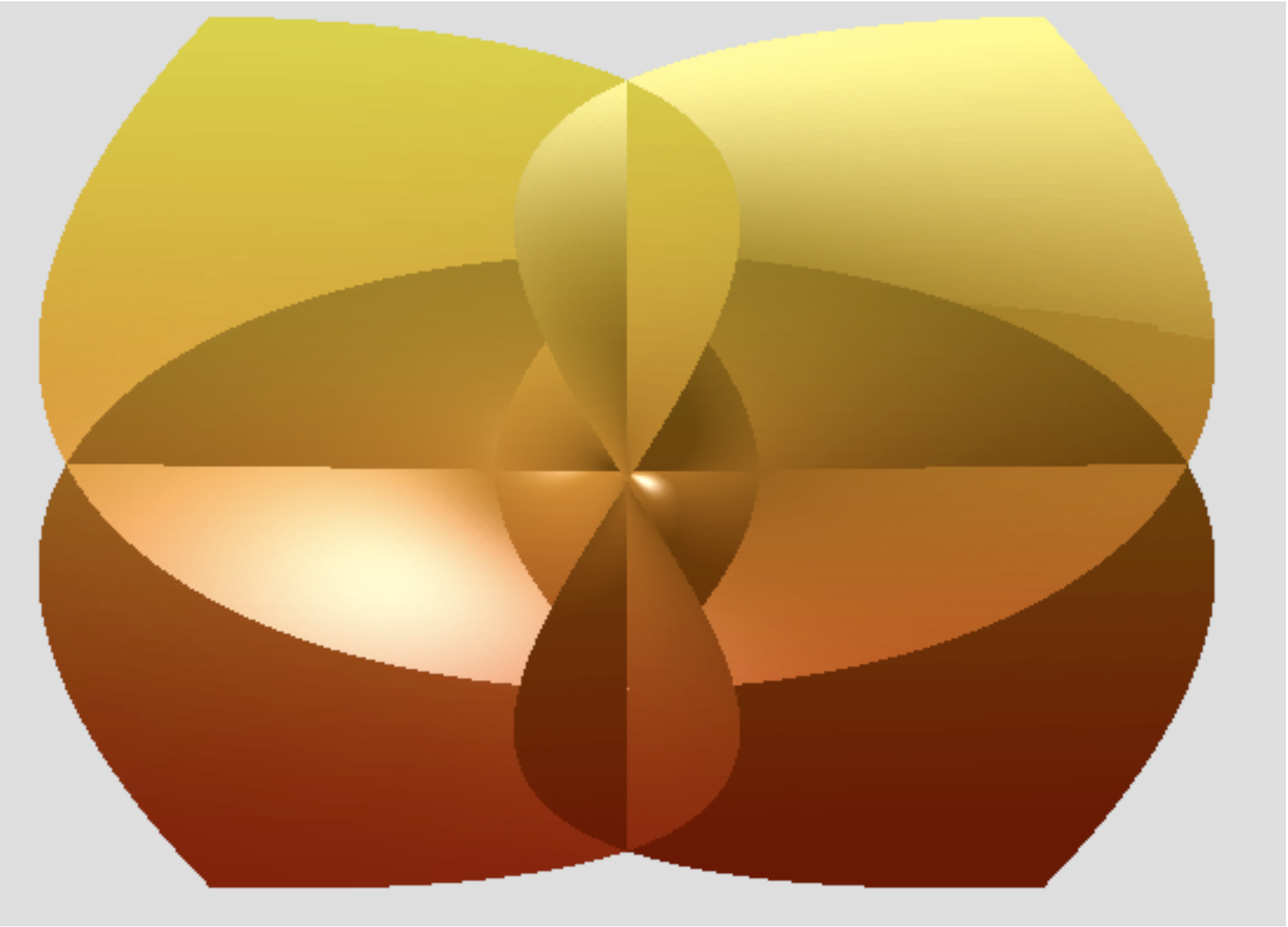}
\\(a) Quartic minimal surface
\end{minipage} \quad
\begin{minipage}[t]{0.6\figwidth}
\centering
\includegraphics[height=1.0\textwidth,width=1.1\textwidth]{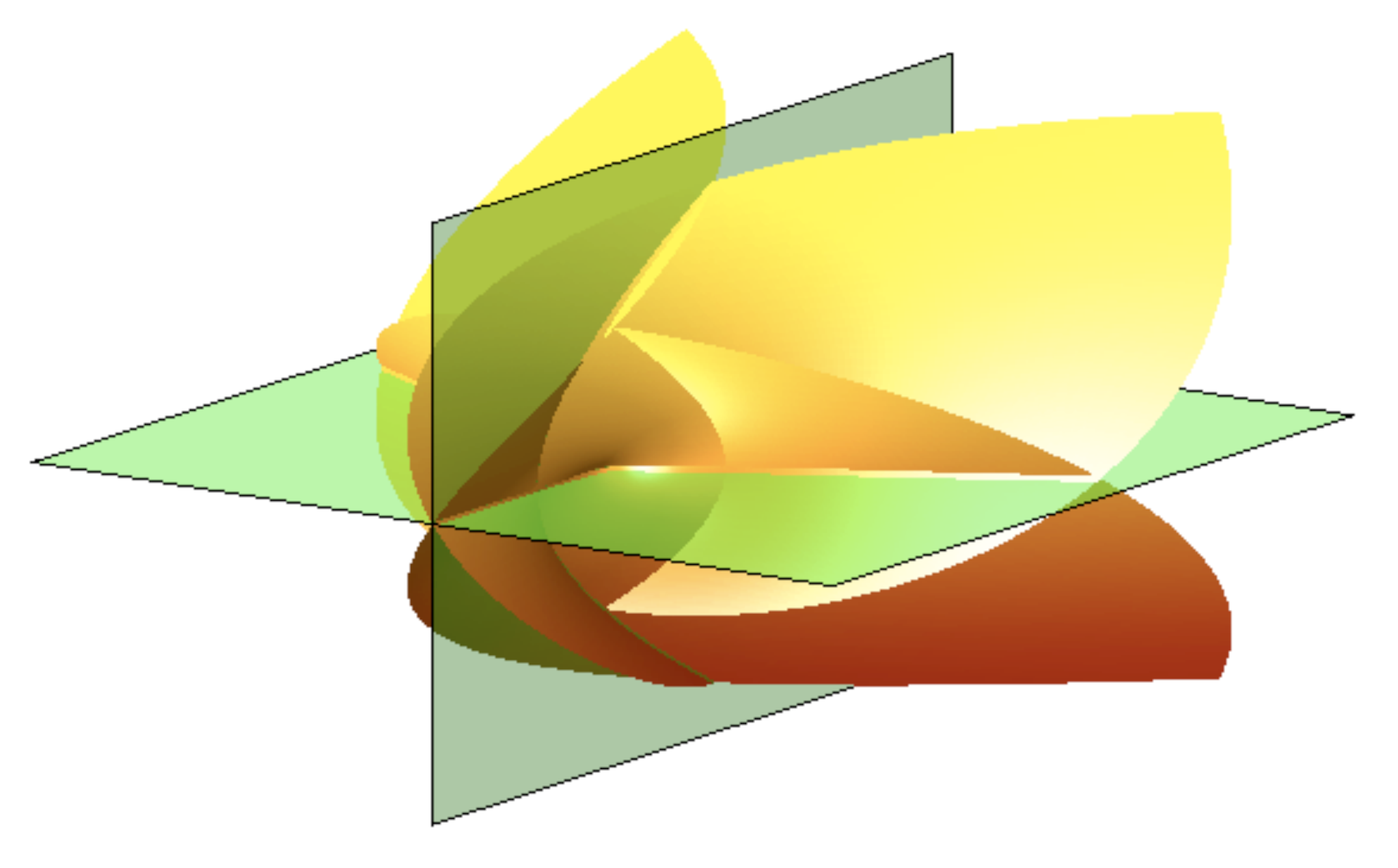}
\\(b) Symmetry plane  
\end{minipage}
\caption{Quartic minimal surface and its symmetry plane.  $\omega=1, [-1,1]\times[-1,1]$.}
\end{figure}

\begin{figure}[t]\label{fig:cubic}
\centering
\begin{minipage}[t]{0.6\figwidth}
\centering
\includegraphics[height=1.0\textwidth,width=1.1\textwidth]{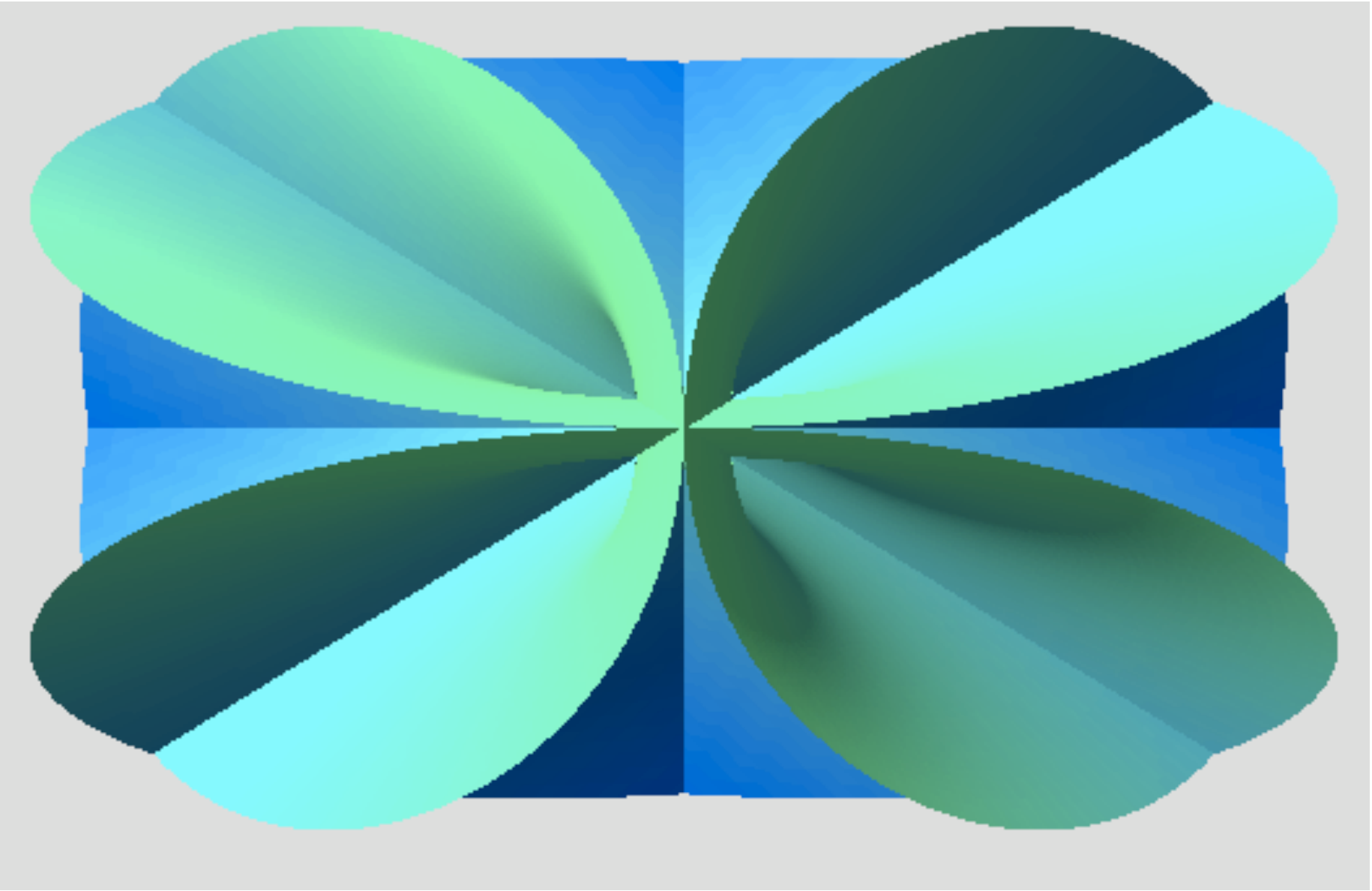}
\\(a) Quintic minimal surface
\end{minipage} \quad
\begin{minipage}[t]{0.6\figwidth}
\centering
\includegraphics[height=1.0\textwidth,width=1.1\textwidth]{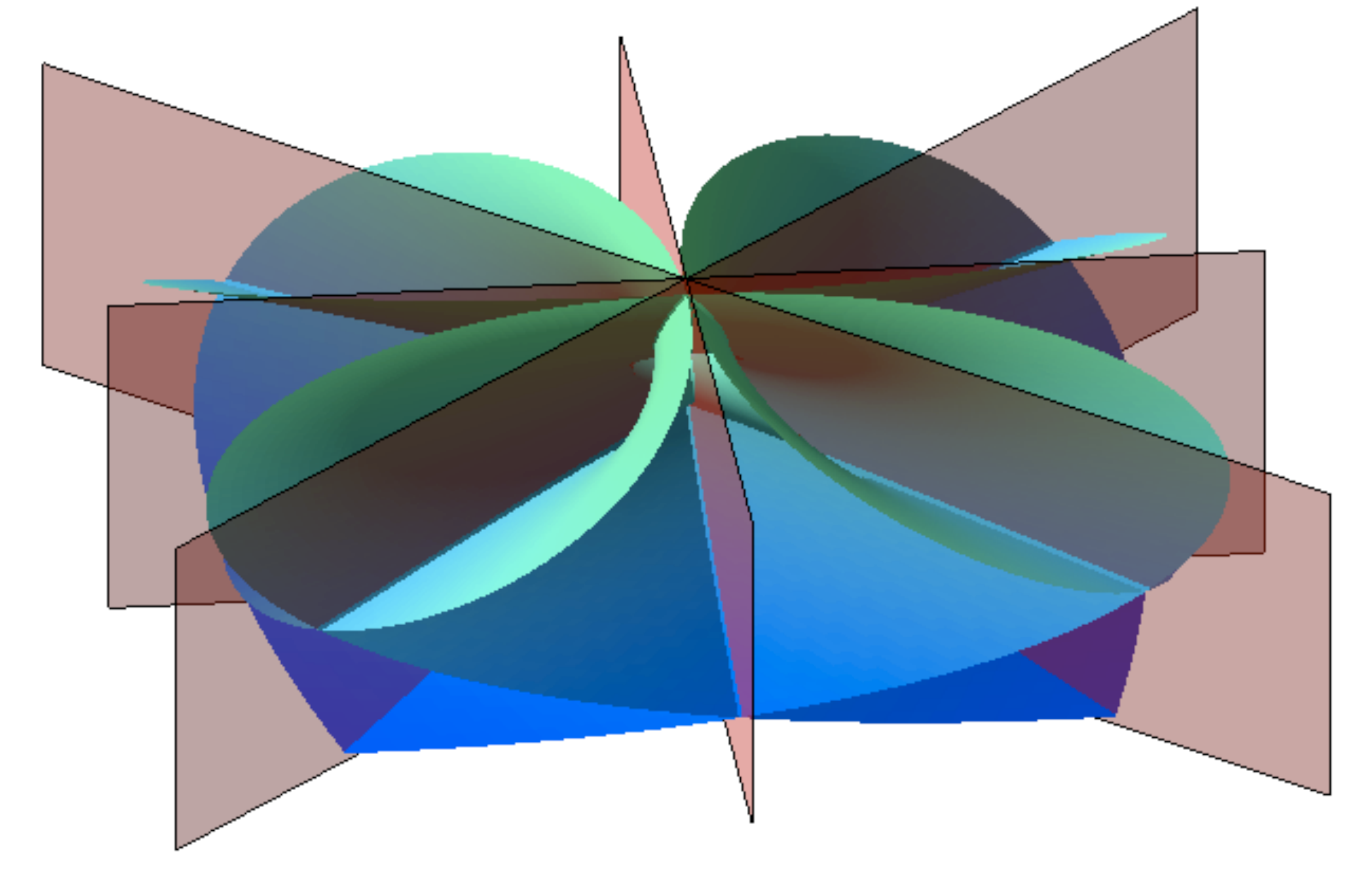}
\\(b) Symmetry plane  
\end{minipage}
\caption{Quintic minimal surface and its symmetry plane.  $\omega=1, [-1,1]\times[-1,1]$.}
\end{figure}

\begin{figure}[t]\label{fig:cubic}
\centering
\begin{minipage}[t]{0.6\figwidth}
\centering
\includegraphics[height=1.0\textwidth,width=1.1\textwidth]{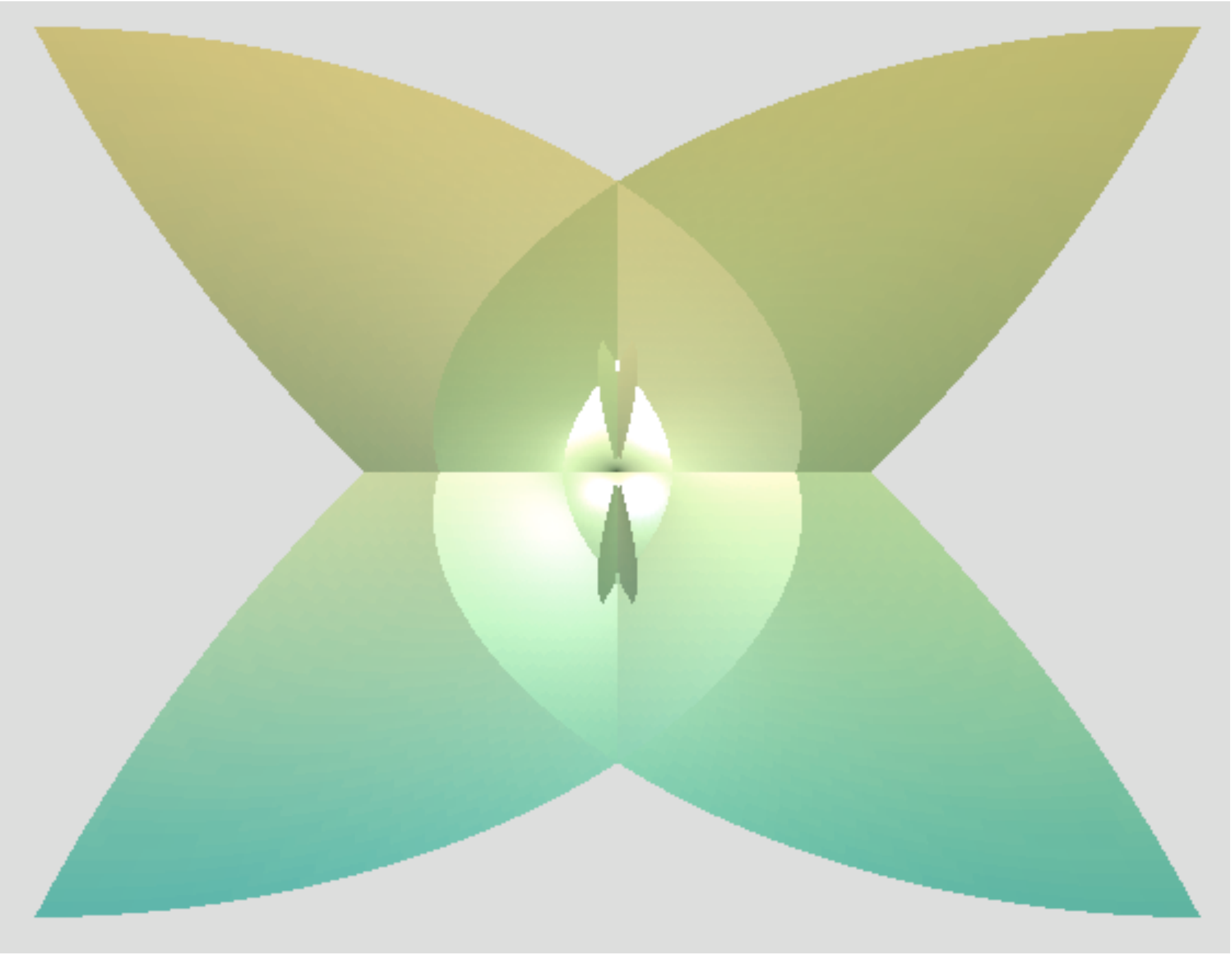}
\\(a) Minimal surface of degree six
\end{minipage} \quad
\begin{minipage}[t]{0.6\figwidth}
\centering
\includegraphics[height=1.0\textwidth,width=1.1\textwidth]{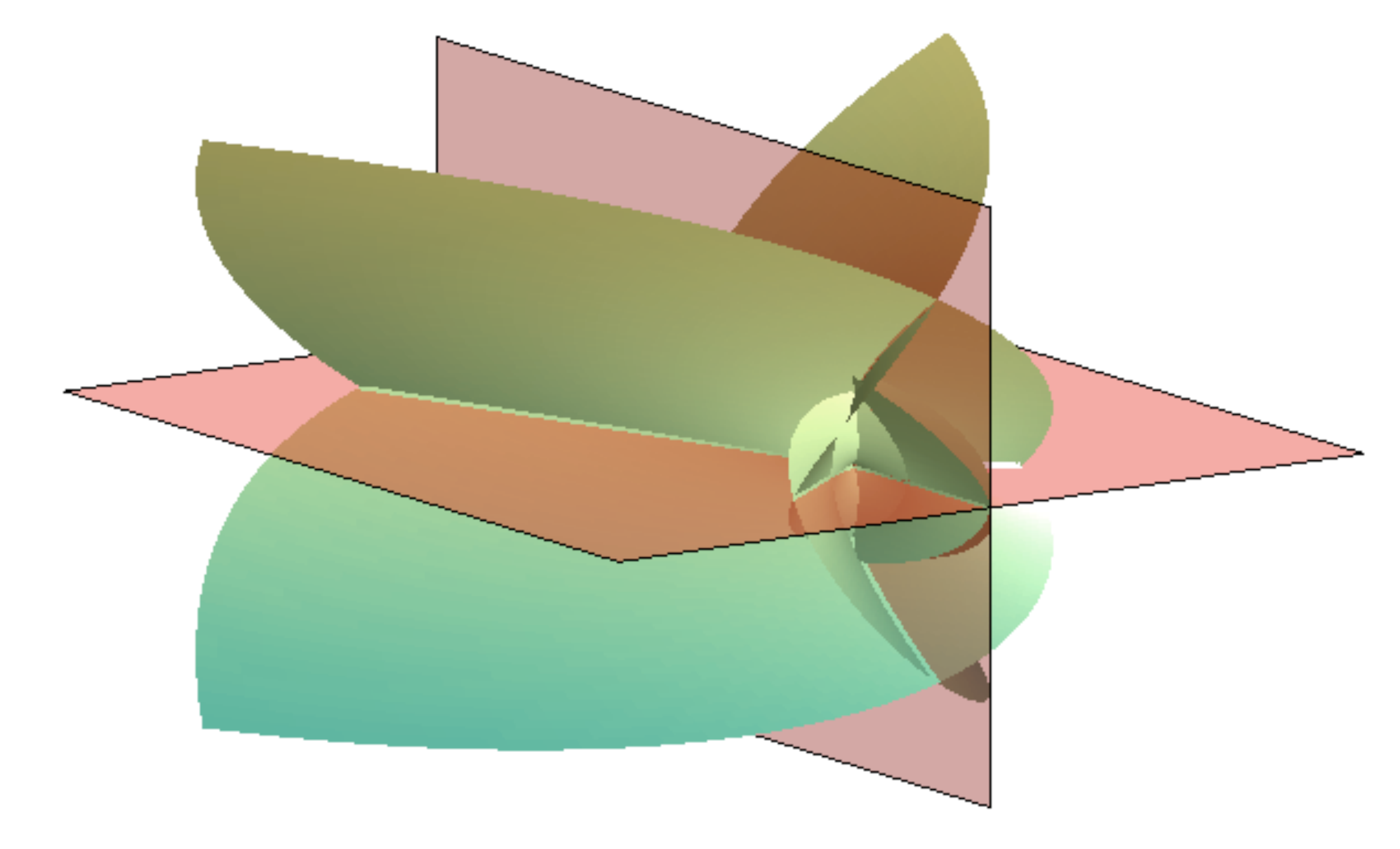}
\\(b) Symmetry plane 
\end{minipage}
\caption{Minimal surface of degree six and its symmetry plane. $\omega=1, [-1,1]\times[-1,1]$.}
\end{figure}

\begin{figure}
\centering
\begin{minipage}[t]{0.5\figwidth}
\centering
\includegraphics[height = 1.0\textwidth,width=1.0\textwidth]{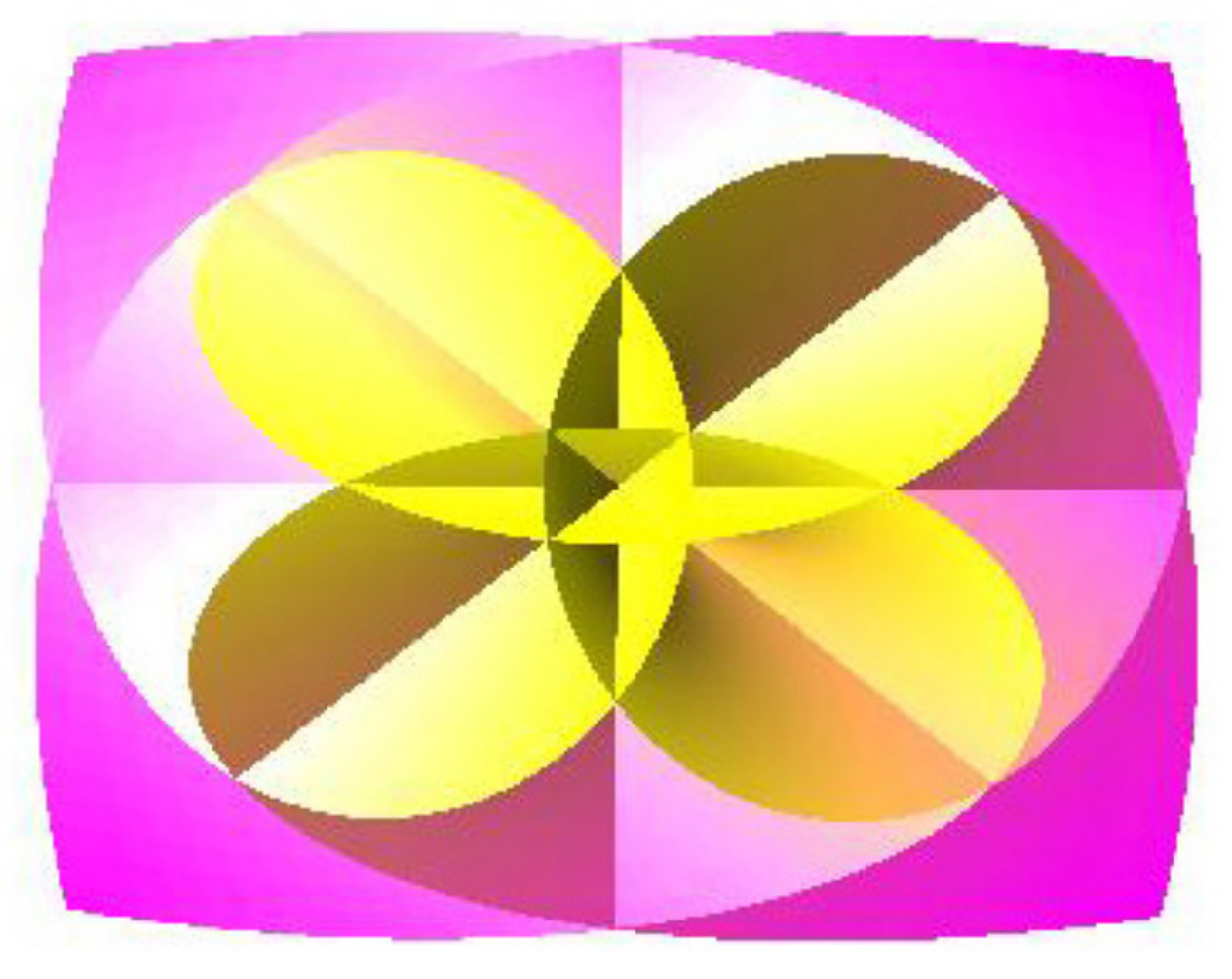}
\\a) $t=0$
\end{minipage}
\begin{minipage}[t]{0.5\figwidth}
\centering
\includegraphics[height = 1.0\textwidth,width=1.0\textwidth]{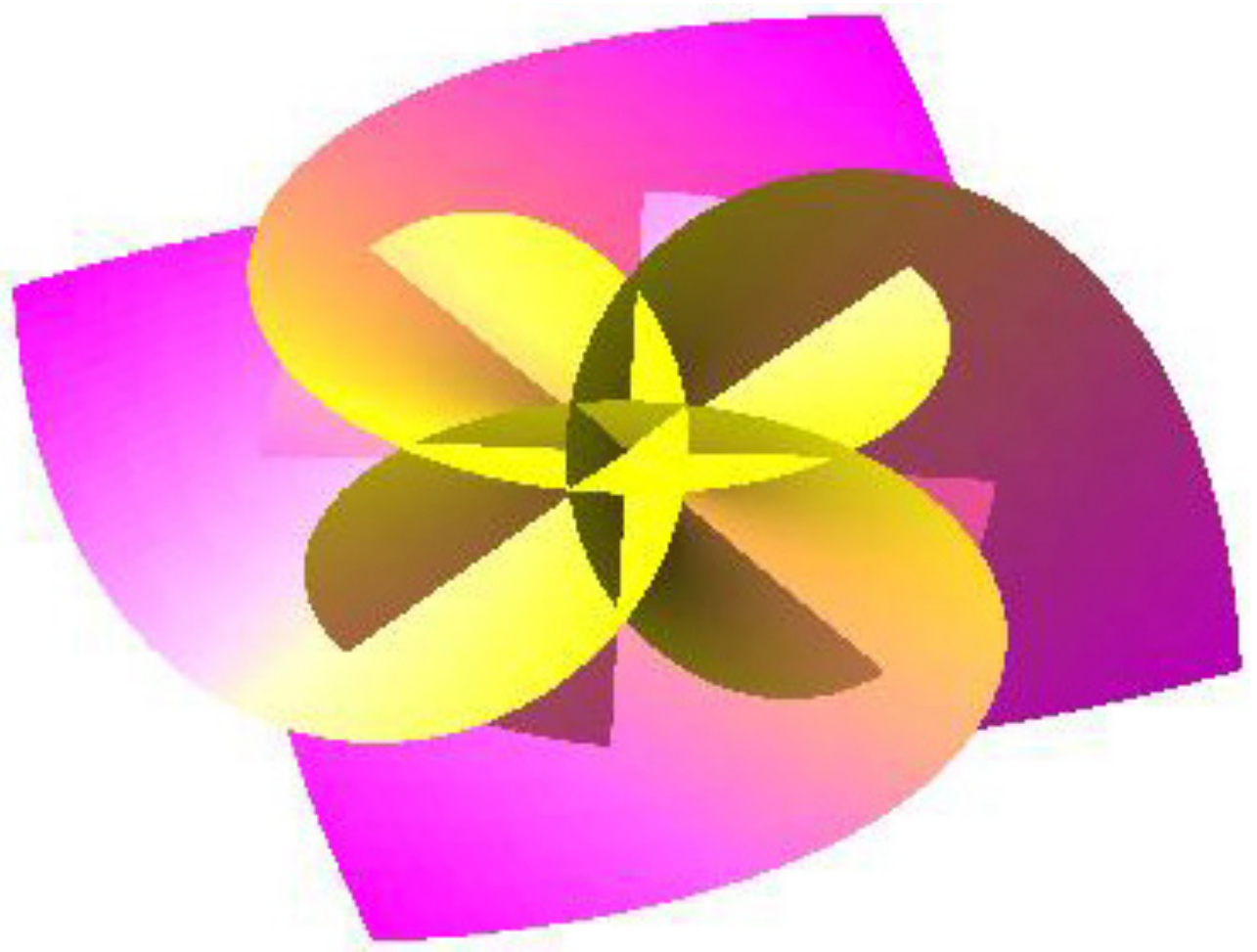}
\\b) $t=\pi/10$
\end{minipage}
\begin{minipage}[t]{0.5\figwidth}
\centering
\includegraphics[height = 1.0\textwidth,width=1.0\textwidth]{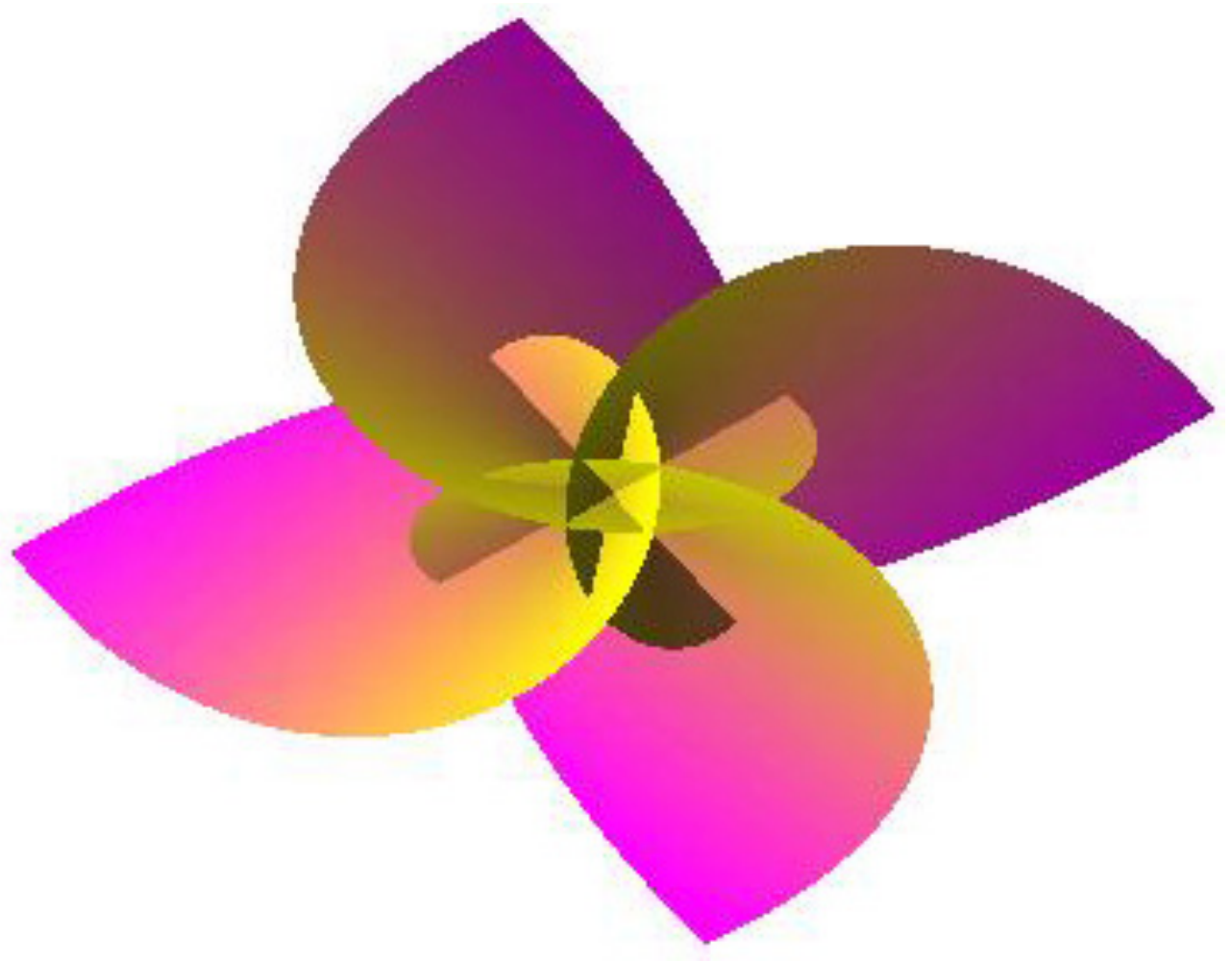}
\\c) $t=\pi/5$
\end{minipage}\\
\begin{minipage}[t]{0.5\figwidth}
\centering
\includegraphics[height = 1.0\textwidth,width=1.0\textwidth]{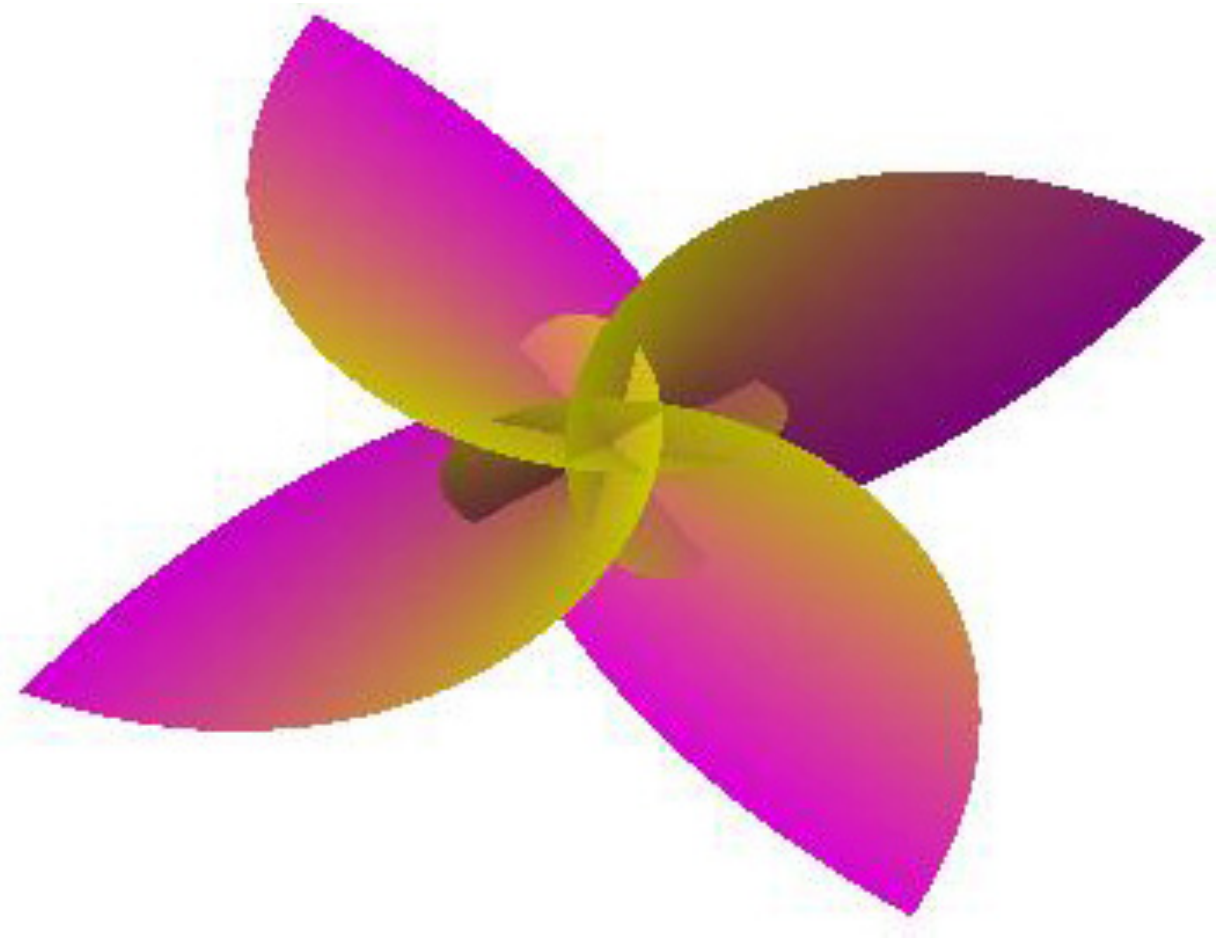}
\\d) $t=3\pi/10$
\end{minipage}
\begin{minipage}[t]{0.5\figwidth}
\centering
\includegraphics[height = 1.0\textwidth,width=1.0\textwidth]{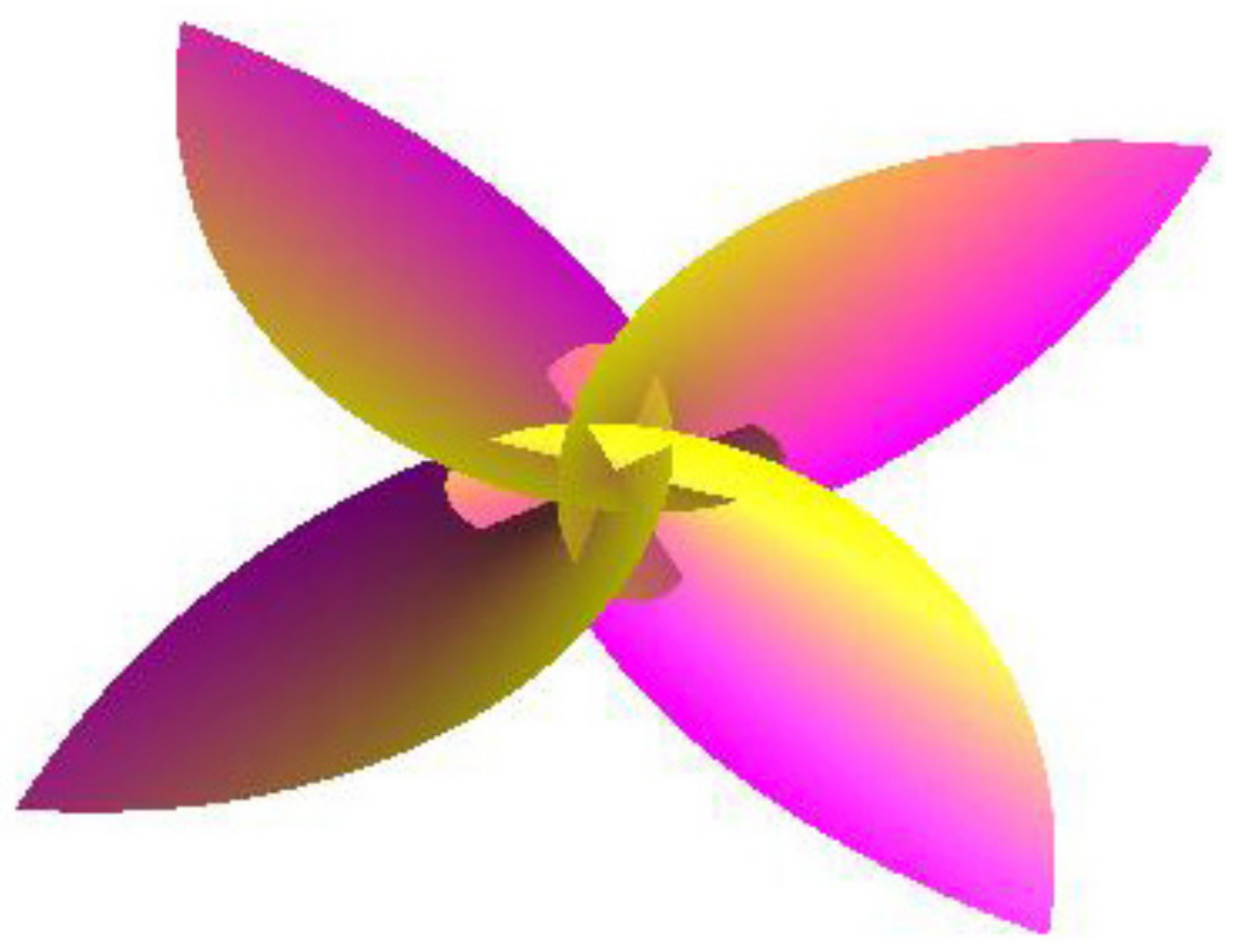}
\\e) $t=2\pi/5$
\end{minipage}
\begin{minipage}[t]{0.5\figwidth}
\centering
\includegraphics[height = 1.0\textwidth,width=1.0\textwidth]{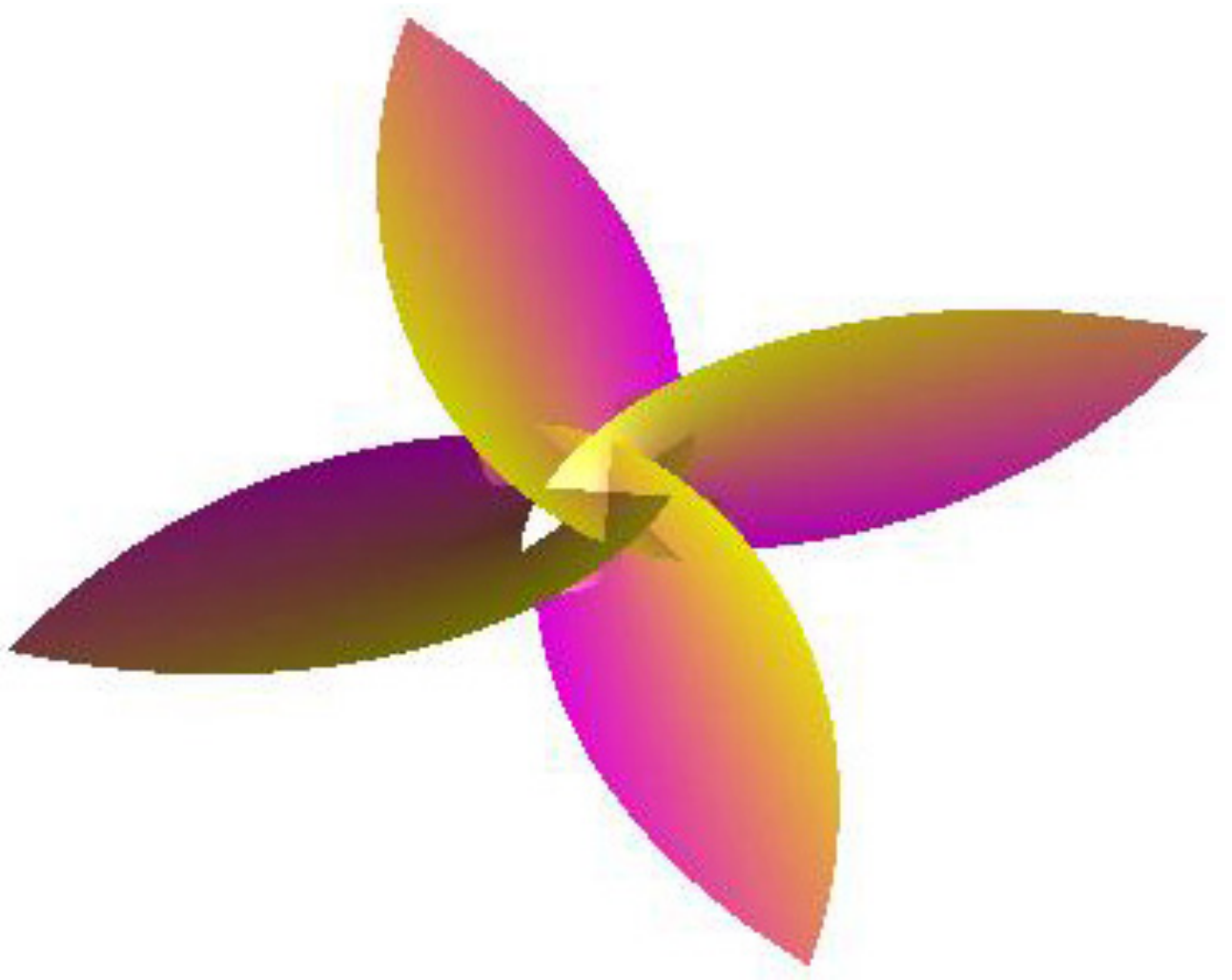}
\\f) $t=\pi/2$
\end{minipage}\\
\caption{Dynamic deformation between
$\emph{\textbf{r}}(u,v)$ and
$\emph{\textbf{s}}(u,v)$, $u,v \in[-4,4]$. }
\label{fig:conju5}
\end{figure}
\section{Conjugate Minimal Surface}
\noindent\textbf{Definition 1. } If two differentiable functions
$p(u,v),q(u,v):  \emph{\textbf{U}} \mapsto \emph{\textbf{R}}$
satisfy the Cauchy-Riemann equations
\begin{equation*}
\frac{\partial p}{\partial u} =  \frac{\partial q}{\partial v},
\frac{\partial p}{\partial v} =  -\frac{\partial q}{\partial u},
\end{equation*}
and both are harmonic, then the functions are said to be
\emph{harmonic conjugate}.

\noindent\textbf{Definition 2. } If $\emph{\textbf{P}}= (p_1, p_2,
p_3)$ and $\emph{\textbf{Q}}=(q_1, q_2, q_3)$ are with isothermal
parameterizations such that $p_k$ and $q_k$ are harmonic conjugate
for $k = 1, 2, 3$, then $\emph{\textbf{P}}$ and $\emph{\textbf{Q}}$
are said to be \emph{parametric conjugate minimal surfaces}.

Helicoid and catenoid are a pair of conjugate minimal surfaces. For
$\textbf{\emph{r}}(u,v)$, we can find out a new pair of conjugate
minimal surfaces as follows.

\noindent\textbf{Theorem 2 }\label{ref:form}
The conjugate minimal surface of $\textbf{r}(u,v)$ has the
following parametric form
\[
\textbf{s} (u,v)=(X_s(u,v), Y_s(u,v), Z_s(u,v))
\]
where
\begin{eqnarray} 
X_s(u,v) & = & -Q_n+\omega Q_{n-2},\nonumber\\
Y_s(u,v) & = & -P_n-\omega P_{n-2},\label{formulacon}\\
Z_s(u,v) & = & \frac{ 2\sqrt {n(n-2)\omega}}{
n-1} Q_{n-1}, \nonumber
\end{eqnarray}
 
It can be proved directly by Lemma \ref{lemma:deriv}. 
From [2],  the surfaces of one-parametric family
\[
\textbf{\emph{C}}_{t}(u,v)=(\cos t)\emph{\textbf{r}}(u,v)+ (\sin t)\emph{\textbf{s}}(u,v)
\] 
\noindent are minimal surfaces with the same first fundamental form.
These minimal surfaces are isometric and have the same Gaussian
curvature at corresponding points. Fig. 5 illustrates the
isometric deformation between 
$\emph{\textbf{r}}(u,v)$ and
$\emph{\textbf{s}}(u,v)$.  It is similar with the isometric
deformation between helicoid and catenoid.
\section{Conclusion}
The explicit parametric formula of polynomial minimal surface 
is presented. It can be considered as the generalization of Enneper 
surface in cubic case. The corresponding properties and classification
of the proposed minimal surface are investigated. The
corresponding conjugate  minimal surface are  constructed and the 
dynamic isometric deformation between them are also implemented. 
   
\noindent\textbf{Acknowlegments }

This work was partially supported  by the National Nature Science
Foundation of China (No.60970079, 60933008), Foundation of State Key Basic Research 973
Development Programming Item of China (No.2004CB318000), and the
Nature Science Foundation of Zhejiang Province, China(No. Y1090718).

\end{document}